\documentclass[fleqn,usenatbib,twocolumn]{mnras}

\usepackage{newtxtext,newtxmath}
\usepackage{xspace}
\usepackage[T1]{fontenc}

\DeclareRobustCommand{\VAN}[3]{#2}
\let\VANthebibliography\thebibliography
\def\thebibliography{\DeclareRobustCommand{\VAN}[3]{##3}\VANthebibliography}

\usepackage{graphicx}	
\usepackage{amsmath}	

\newcommand\asloth{\textsc{a-sloth}\xspace}

\newcommand{\mum}{$\rm \mu \mathrm{m}~$}

\newcommand\rev[1]{\textcolor{black}{#1}}
\newcommand\revrev[1]{\textcolor{black}{#1}}

\newcommand{\OIII}{[\textsc{Oiii}]\,}

\title[Dust in High-redshift Galaxies]{A Photon Burst Clears the Earliest Dusty Galaxies: Modelling Dust in High-redshift Galaxies from ALMA to JWST}

 \author[Tsuna et al.]{
 Daichi Tsuna$^{1,2}$\thanks{Contact e-mail: \href{mailto:tsuna@caltech.edu}{tsuna@caltech.edu}},
 Yurina Nakazato$^{3}$,
 and  Tilman Hartwig$^{3,4,5}$
 \\
$^{1}$ TAPIR, Mailcode 350-17, California Institute of Technology, Pasadena, CA 91125, USA \\
$^{2}$ Research Center for the Early Universe (RESCEU), School of Science, The University of Tokyo, Bunkyo, Tokyo 113-0033, Japan \\
$^{3}$ Department of Physics, School of Science, The University of Tokyo, Bunkyo, Tokyo 113-0033, Japan \\
$^{4}$ Institute for Physics of Intelligence, School of Science, The University of Tokyo, Bunkyo, Tokyo 113-0033, Japan\\
$^{5}$ Kavli Institute for the Physics and Mathematics of the Universe (WPI), The University of Tokyo, Kashiwa, Chiba 277-8583, Japan
}

\begin{document}
\label{firstpage}
\pagerange{\pageref{firstpage}--\pageref{lastpage}}
\maketitle

 \begin{abstract}
The generation and evolution of dust in galaxies are important tracers for star formation, and can characterize the rest-frame ultraviolet to infrared emission from the galaxies. In particular understanding dust in high-redshift galaxies are important for observational cosmology, as they would be necessary to extract information on star formation in the early universe. We update the public semi-analytical model \asloth (Ancient Stars and Local Observables by Tracing Halos) to model the evolution of dust, focusing on high-redshift \rev{star-forming} galaxies \rev{with stellar masses of $\sim 10^8$--$10^{10}M_\odot$} observed by ALMA ($z\approx 7$) and JWST ($z\approx 11$). We find that these galaxies should qualitatively differ in their star formation properties; while the samples in ALMA are explained by dust growth in normal star-forming galaxies, the lack of dust in the samples by JWST requires dust ejection by radiation pressure due to recent highly efficient star-formation \rev{within a few 10 Myr}, with order 100 times higher efficiency than normal galaxies calibrated by \asloth. Depending on where the JWST galaxies locate on the luminosity function, \rev{their bursty star formation histories inferred from our model} can have impacts for rates of star formation, \rev{supernova} explosion, stellar feedback, \rev{and detectability of dusty, mature galaxies} in the very early universe. 
\end{abstract}

 \begin{keywords}
methods: numerical -- galaxies: high-redshift -- (ISM:) dust, extinction
 \end{keywords}



\section{Introduction}

Dust is an important component in galaxies that \rev{shapes} the emission from them in a wide range of wavelengths. Dust generally absorbs the ultraviolet and optical light, and re-emits them in the infrared \citep[e.g.,][]{Spitzer78,Bruzual88,Witt92}. For example dust in our Galaxy, with mass of $\sim10^8\ M_\odot$ distributed throughout the disc \citep{Planck11}, has prevented a large fraction of historical Galactic supernovae (SNe) from being observed by human eye \citep[e.g.,][]{vandenBergh75,Tammann94,Adams13}.

Studying dust in high-redshift galaxies in these wavelengths is a direct probe of the environment of galaxies in its youth. \rev{Star-forming galaxies at redshifts of $z=4$--$7$ probed by the Atacama Large Millimeter Array (ALMA) find a dusty environment \citep[e.g.,][]{Watson15,Knudsen17,Laporte17,Hashimoto19,Tamura19,Bouwens22,dayal22,Inami22}. Though estimates of dust masses can be sensitive to the uncertain dust temperature \citep[e.g.,][]{Sommovigo20}, these galaxies are generally found to have dust to stellar mass ratios of order $0.1$--$1$ per cent, and in some case close to 10 per cent \citep{Marrone18}}. Recent observations by the James Webb Space Telescope (JWST) instead have discovered galaxy candidates at $z\gtrsim 10$ that are nearly dust-free with extremely low dust attenuation \citep[e.g.,][]{Naidu22,Finkelstein22,Castellano22,Donnan23,Morishita23,Atek23,Furtak23,Harikane23}, as well as dust-obscured ones \citep{Rodighiero23}. While a definitive conclusion may not yet be possible, these results may already indicate a complex growth history of dust in the early universe.

On the theoretical side, it would be important to obtain the dust properties of a large number of galaxies at high-redshift that enable comparison with observations. Such modelling has been done in the past decade \citep[e.g,][]{Dwek07,Valiante09,Valiante11,deBennassuti14,Valiante14,Mancini15,Gioannini17,McKinnon17,Popping17,Aoyama18,Vijayan19,Graziani20,Triani20,dayal22,Ferrara22b,DiCesare23,Mauerhofer23}, with many of them aiming to understand the dust content of galaxies observed by ALMA. As the number of JWST galaxies is expected to grow, a statistical argument on the processes that govern dust evolution \rev{at high redshifts} can become possible \rev{(see e.g., \citealt{Witstok23} for discussion on a galaxy JADES-GS-z6-0)}. In addition, the extended redshift range of JWST may probe specific physics for star formation at zero and low metallicities.

Various processes can create, grow, or destroy dust. The dust is generally created in the ejected material from stars, and \rev{grows} through its accretion of metals in the interstellar medium \citep[e.g.,][]{Liffman89,Dwek98,Draine09,Asano13}. Stellar feedback can act against the growth of dust, the representative being destruction and ejection due to SN explosions \citep[e.g.,][]{McKee89}. Recently it has been pointed out that radiation driven ejection of dust (and gas) may be important in star-forming galaxies at high redshift observed \rev{with} JWST \citep{Ferrara23,Ziparo23,Fiore23}. While these works analytically demonstrate the importance of ejection by radiation pressure, they do not conduct a detailed time-dependent modelling of the dust evolution. 

In this work we study the evolution of dust in these high-redshift galaxies using the semi-analytical model \asloth (Ancient Stars and Local Observables by Tracing Halos; \citealt{Hartwig22,Magg22,Chen22}), taking into account the \rev{aforementioned physics of dust production, growth and destruction/ejection} as much and detailed as a semi-analytical model allows. Such semi-analytical approach allows modelling time-dependent dust production, growth, and ejection for a large number of galaxies at a given redshift. Furthermore, our code, which is publicly available as an update to \asloth \rev{(upon acceptance of this paper)}, would \rev{enable the user to} update the dust physics as desired.

The structure of this paper is as follows. In Section \ref{sec:Methods} we overview the \asloth code, and describe our methodology for newly implementing the formation and evolution of dust in this code. In Sections \ref{sec:results} and \ref{sec:jwst} we present our results, and discuss this in the context of recent observational and theoretical findings. We argue in Section \ref{sec:jwst} of a requirement of efficient star formation to explain the deficiency of dust. We conclude in Section \ref{sec:conclusion}, \rev{by mentioning implications for future observations} and listing some of the future avenues to refine our model.

\section{Methods}
\label{sec:Methods}
We use the public semi-analytical model \asloth\footnote{\url{https://gitlab.com/thartwig/asloth}} to simulate the formation and evolution of galaxies. The code is based on dark matter merger trees, and includes baryonic physics in a semi-analytical way that realizes a large number of galaxy samples with high temporal and mass resolution. The input parameters for each galaxy are the \rev{final} redshift $z$, and the halo mass $M_{\rm halo, f}$ at this final redshift. From this parent halo, \asloth first generates the merger tree backwards in time \rev{from redshift $z$,} and then simulates the baryonic physics forward in time \rev{along the tree down to redshift $z$}. We consider a wide range of halo masses of $M_{\rm halo, f}=10^8$--$10^{12}\ M_\odot$ \rev{to be assembled at the final redshifts $z=7$ and $11$.} \rev{This mass range} covers the most relevant parameter range for high-z galaxies detected with ALMA and JWST. The halo mass is sampled by 0.1 dex intervals, and for each $(M_{\rm halo,f},z)$ pair we consider 10 random realizations of the merger tree.

\asloth is ideally suited to study dust in high-z galaxies because it is the only public model that resolves individual stars and minihalos at all redshift. It is well calibrated \citep{Hartwig22,uysal23} and has an explicit formulation for Pop~III star formation, which is important for high-z galaxies \citep{riaz22}. Moreover, \asloth allows us to explore a large parameter range (in halo masses and redshift), so that we can eventually compare and select the models that best reproduce observations.

\subsection{Baryonic Physics in A-SLOTH}
For each merger tree branch, \asloth follows the evolution of the hot ISM gas, cold ISM gas, and stars. \rev{In the code the cold ISM corresponds to the cold neutral medium, whereas the hot ISM corresponds to everything else, including the warm gas and hot ionized gas \citep[see e.g.,][]{McKee77,Klessen16}}. As in most semi-analytical models these are one-zone, with each component defined a representative spatial scale and mass (see also Figure 1 of \citealt{Chen22}). The hot gas is assumed to span over a virial radius $R_{\rm vir}$. On the other hand, the cold gas is assumed to occupy a more compact region of radii
\begin{equation}
    R_{\rm s}\equiv \frac{R_{\rm vir}(z, M_{\rm halo})}{c_{\rm dm}(z, M_{\rm halo})},
    \label{eq:R_s}
\end{equation}
where $R_{\rm s}$ is the scale radius of the dark matter profile \citep{Navarro97}, and $c_{\rm dm}(>1)$ is the concentration factor of the dark matter halo with a constant value of $3.5$ at $z>10$ and a function of halo mass and redshift at $z<10$ calibrated by cosmological simulations (\citealt{Correa15}; see also Appendix A of \citealt{Hartwig22}). The stars co-exist with the cold gas, but they may in fact be even more compact than the radii in equation (\ref{eq:R_s}) given the clumpy nature of star-forming regions. While the extent of the stellar population is irrelevant in the original version of \asloth, it can be important when dust is considered, and we discuss this in detail in Section \ref{sec:rad_feedback}.

At each adaptive timestep $\delta t_i$, we update the masses of the cold ISM $M_{\rm cold}$, hot ISM $M_{\rm hot}$, stars $M_*$, and the outflow by feedback processes $M_{\rm out}$ by the following equations \citep{Hartwig22}
\begin{align}
    M_{\rm cold}^{i+1} &= M_{\rm cold}^{i} + \frac{\delta t_i M_{\rm hot}^{i}}{t^{i}_{\rm dyn}} - \delta M_{\rm out, cold}^{i} \nonumber \label{eq:coldgas}\\
    &- \delta M_{\rm heat}^{i} - \eta_*M^{i}_{\rm cold}\frac{\delta t_i}{t^{i}_{\rm cold,ff}} - \frac{\delta t_i M^{i}_{\rm cold}}{t_{\rm rad, cold}}, \\
    M_{\rm hot}^{i+1} &= M_{\rm hot}^{i} - \frac{\delta t_i M_{\rm hot}^{i}}{t^{i}_{\rm dyn}} - \delta M_{\rm out, hot}^{i} \nonumber \\
    &+ \delta M_{\rm heat}^{i} + \delta M_{\rm acc, hot}^{i} - \frac{\delta t_i M^{i}_{\rm hot}}{t_{\rm rad, hot}} , \\
    M_*^{i+1} &= M_*^{i}+\eta_*M^{i}_{\rm cold}\frac{\delta t_i}{t^{i}_{\rm cold,ff}},\\
    M_{\rm out}^{i+1} &= M_{\rm out}^{i} + \delta M_{\rm out, cold}^{i} + \delta M_{\rm out, hot}^{i},
\end{align}
where $t_{\rm dyn}$ is the dynamical timescale of the central region of the halo, $t_{\rm cold,ff}$ is the free-fall timescale of the cold gas, \rev{$\delta M_{\rm heat}$ is the mass of cold gas converted to hot gas by ionizing photons, and $\delta M_{\rm acc, hot}$ is the mass of hot gas accreted from the intergalactic medium (IGM). Finally, $t_{\rm rad, hot}, t_{\rm rad, cold}$ are timescales newly incorporated from \cite{Hartwig22} regarding ejection of gas and dust driven by radiation from stars, and are derived in Section \ref{sec:rad_feedback}.}

\rev{Both Pop-II and Pop-III star formation are taken into account with an analytical metallicity criterion, and the code adopts different initial mass functions (IMFs),  metal yields, and dimensionless star-formation efficiencies $\eta_*$ (respectively defined as $\eta_{\rm II}, \eta_{\rm III}$ in the rest of the paper) for each population.} The code keeps track of individual stars, with appropriate delay times between formation and core-collapse for massive stars. \rev{The $\eta_*$ in \asloth is different from the usual definition of ``star-formation efficiency", which is the (time-integrated) fraction of gas that are converted into stars. In \asloth the definition is based on timescale, i.e. how many free-fall times $t^{i}_{\rm cold, ff}=(G\rho^{i}_{\rm cold})^{-1/2}$, defined for cold gas with uniform density $\rho^{i}_{\rm cold}=M^{i}_{\rm cold}/(4\pi R_{\rm s}^3/3)$ and $G$ being the gravitational constant, it takes for cold gas to convert into stars. If the cold gas has local overdensities (such as clumps) the value of $\eta_*$ can be much greater and even exceed unity, as will be considered later in Section \ref{sec:jwst}.} 

\rev{In addition to $\eta_{\rm II}$ and $\eta_{\rm III}$, \asloth has other free parameters such as the maximum mass of Pop-III stars $M_{\rm max}$, outflow efficiency (slope $\alpha_{\rm out}$ and normalization $M_{\rm out, norm}$ of virial mass-dependent efficiency) and the escape fraction of ionizing photons from Pop-II stars $f_{\rm esc, II}$. In our default calculations (Section \ref{sec:results}) their values are fixed to those calibrated from six local and redshift-dependent observables, as $\eta_{\rm II}=0.19$, $\eta_{\rm III}=0.38$, $M_{\rm max}=210\ M_\odot$, $\alpha_{\rm out}=0.86$, $M_{\rm out, norm}=7.5\times 10^9\ M_\odot$ and $f_{\rm esc, II}=0.37$ (see Section 3 and Table 3 of \citealt{Hartwig22}).}

We refer the details on the derivation of each term in the governing equations to the methods paper of \cite{Hartwig22}, but for completeness we list the essential physics with corresponding sections in \cite{Hartwig22}:
\begin{itemize}
    \item The hot gas cools and compresses over the dynamical timescale, converting hot gas into cold gas (Sec 2.2).
    \item The cold gas further compresses and form stars, which some of them explode as SNe after their lifetime (Sec 2.3.1).
    \item Hot and cold gas can be ejected out of the halo depending on their binding energies relative to the injected energy from SNe (Sec 2.3.4).
    \item The ionizing photons from the stars heat the surrounding cold ISM, converting cold gas into hot gas (Sec 2.3.3).
    \item The halo accretes additional gas from the IGM, which is fed only to the hot ISM (Sec 2.2).
\end{itemize}

\subsection{Dust Evolution Model}
To track the evolution of dust, we newly incorporate in \asloth the following set of equations following the prescriptions of \cite{deBennassuti14} and \cite{Valiante14}
\begin{align}    
    M_{\rm d, cold}^{i+1} =& M_{\rm d, cold}^{i}+ \frac{\delta t_i M_{\rm d, hot}^{i}}{t_{\rm dyn}} - (\delta M_{\rm out, cold}^{i}+\delta M_{\rm heat}^{i}+ \delta M_*^{i})\frac{M_{\rm d, cold}^{i}}{M_{\rm cold}^{i}} \nonumber \\
    &+ \frac{\delta t_i M_{\rm d, cold}^{i}}{t_{\rm grow}} - \frac{\delta t_i M_{\rm d, cold}^{i}}{t_{\rm rad, cold}}, \\
    M_{\rm d, hot}^{i+1} =& M_{\rm d, hot}^{i} - \frac{\delta t_i M_{\rm d, hot}^{i}}{t_{\rm dyn}} - (\delta M_{\rm out, hot}^{i}-\delta M_{\rm heat}^{i} )\frac{M_{\rm d, hot}^{i}}{M_{\rm hot}^{i}} \nonumber \\
    &+ \sum_{\rm SN} (Y^{i}_{\rm d} - \delta M^{i}_{\rm dest}) - \frac{\delta t_i M_{\rm d, hot}^{i}}{t_{\rm rad, hot}},
\end{align}
where the second and third terms of each equation are the mass exchange among cold ISM, hot ISM and stars, multiplied by the corresponding dust fraction in each phase. Here it is implicitly assumed that dust and gas are dynamically coupled; for a dust particle of size $\bar{a}=0.1\,\mu$m and mass $m_{\rm d}\approx 10^{-14}$ g (for a typical density of $\approx 2$--$3$ g cm$^{-3}$; \citealt{Laor93}), the coupling timescale $t_{\rm coup}$ given by momentum exchange with gas of number density $n_{\rm gas}$ and mass of proton mass $m_{\rm p}$ is
\begin{align}
    t_{\rm coup} \approx & \left[(\pi \bar{a}^2)n_{\rm gas}\Delta v\frac{m_{\rm p}}{m_{\rm d}}\right]^{-1} \nonumber \\
    \approx & 0.6\ {\rm Myr}\left(\frac{n_{\rm gas}}{1\ {\rm cm^{-3}}}\right)^{-1}\left(\frac{\Delta v}{10\ {\rm km\ s^{-1}}}\right)^{-1} \left(\frac{\bar{a}}{0.1\,\mu \rm m}\right)^2 \left(\frac{m_{\rm d}}{10^{-14}{\,\rm g}}\right)^{-1},
\end{align}
where $\Delta v$ is the typical relative velocity between the particles, which is governed by the thermal motion of the gas. This is generally much shorter compared with the dynamical (free-fall) timescale of the gas
\begin{align}
    t_{\rm dyn}\approx& (Gn_{\rm gas}m_{\rm p})^{-1/2}\sim 100\ {\rm Myr}\left(\frac{n_{\rm gas}}{1\ {\rm cm^{-3}}}\right)^{-1/2},
\end{align}
for cold ISM, and likely also for hot ISM. If the dust is in the vicinity of stars it can be charged by stellar radiation, which further enhance the coupling due to Coulomb collisions and magnetic fields \citep{Draine79,Murray05}. 

For dust in cold ISM, we consider its growth by accretion of metals with a timescale \citep{Asano13}
\begin{align}
    t_{\rm grow} =& 30\ {\rm Myr}\left(\frac{\bar{a}}{0.1{\rm \mu m}}\right)\left(\frac{n_{\rm cold}}{100\ {\rm cm^{-3}}}\right)^{-1} \left(\frac{T_{\rm cold}}{50\ {\rm K}}\right)^{-1/2} \nonumber \\
    &\times \left(\frac{Z}{Z_\odot}\right)^{-1} \left(1-\frac{M_{\rm d, cold}^{i}}{M_{\rm metal}^{i}}\right)^{-1} \nonumber \\
    \equiv& t_{\rm acc,\odot} \left(\frac{Z}{Z_\odot}\right)^{-1} \left(1-\frac{M_{\rm d, cold}^{i}}{M_{\rm metal}^{i}}\right)^{-1},
    \label{eq:t_grow}
\end{align}
where $\bar{a}$ is the grain size, $n_{\rm cold}$ and $T_{\rm cold}$ are the density and temperature in the cold ISM, and $Z$ is the metallicity (with solar value of $Z_\odot=0.0134$; \citealt{Asplund09}). The last factor is the fraction of metals that are in the gas phase, which sets the metal budget available for accretion. The values of $\bar{a}, n_{\rm cold}, T_{\rm cold}$ is uncertain, and would likely vary from one galaxy to another. In this work, we adopt two representative cases of $t_{\rm acc,\odot}=30$ and $3\ {\rm Myr}$, which cover the values inferred in past modelling of high-redshift galaxies \citep[e.g.,][]{deBennassuti14,Mancini15,dayal22}.  

For the dust in the hot ISM, we consider the dust yield from SNe and dust destruction by the SN shocks. Since each SN event is resolved in \asloth at each timestep, we add the contribution from each SN within a given timestep. 

The SN dust yield $Y^{i}_{\rm d}$ is uncertain, with various predictions for works with different treatments of the explosion and dust physics \citep[e.g.][]{Kozasa89,Todini01,Nozawa03,Schneider04,Bianchi07,Nozawa07,Cherchneff09,Cherchneff10,Marassi15,Marassi19}, and also with some degree of stochasticity in the mass-metallicity parameter space. The yield can also be altered by binary evolution, which can change the mass and structure of massive stars. In this work we remain agnostic to a particular result, and randomly assign a dust yield within a log-uniform distribution of [$0.03\ M_\odot, 1.1\ M_\odot$], where the limits are from a broad range of dust masses inferred from local SN remnants and pulsar-wind nebulae \citep{Matsuura15,Temim17,Rho18,Priestley20,Niculescu21}. \rev{The above observations probe cold dust that has not encountered the reverse shock of the SN, and a significant fraction of the dust can be destructed at the later stages of the SN remnant phase. We thus assume that 30 per cent of the dust can survive and be injected to the ISM, a value motivated from recent studies \citep{Kirchschlager19,Slavin20}. The survival fraction can depend on the explosion energy, as well as properties intrinsic to the dust such as the grain size distribution and composition \citep[e.g.][]{Nozawa07,Bianchi07,Silvia10,Micelotta16,Biscaro16,Kirchschlager19,Slavin20,Kirchschlager23}.} For pair-instability SNe where confident observations are absent (see \citealt{Schulze23} for a recent candidate), we assume that each SN produces $Y^{i}_{\rm d}=3\ M_\odot$ \citep{Nozawa07,Nozawa03,Schneider04}. This takes into account the {\it in-situ} destruction of dust by the reverse shock in the SN remnant phase, here assumed for a typical ISM density in the Milky Way of $0.1$--$1\ {\rm cm^{-3}}$.

The dust destruction by shocks from SN in the Sedov-Taylor stage is formulated as \citep{McKee89}
\begin{align}
    \delta M_{\rm dest}^{i} &= \frac{M_{\rm d, hot}^{i}}{M_{\rm hot}^{i}}\epsilon_{\rm d}M_{\rm swept} \nonumber \\
    &\approx  \frac{\epsilon_{\rm d} M_{\rm d, hot}^{i}}{M_{\rm hot}^{i}}(1700\ M_\odot)f_{\rm SN}\left(\frac{E_{\rm SN}}{10^{51}\ {\rm erg}}\right) \left(\frac{v_{\rm crit}}{200\ {\rm km\ s^{-1}}}\right)^{-2},
\end{align}
where $\epsilon_{\rm d}$ is the dust destruction efficiency, $f_{\rm SN}$ is a reduction factor that takes into account the fact that progenitors of some of the SNe are clustered, $E_{\rm SN}$ is the SN explosion energy, $v_{\rm crit}\sim 200\ {\rm km\ s^{-1}}$ is the SN shock velocity at the transition from the (energy-conserving) Sedov-Taylor phase to the snowplow phase, \rev{and $M_{\rm swept}$ is the mass of the swept-up ISM \revrev{shock-accelerated} to a velocity of at least $v_{\rm crit}$}. Here we follow \cite{deBennassuti14} and adopt $\epsilon_{\rm d}=0.48, f_{\rm SN}=0.15$ for core-collapse SNe \citep{Jones96}, and $\epsilon_{\rm d}=0.6, f_{\rm SN}=1$ for pair-instability SNe \citep{Nozawa06}. The explosion energies are set to the default values in \asloth, with $10^{51}$ erg for core-collapse SNe and $3.3\times 10^{52}$ erg for pair-instability SNe.

\subsection{Implementation of Radiation Feedback}
\label{sec:rad_feedback}

\begin{figure}
 \centering
\includegraphics[width=\linewidth]{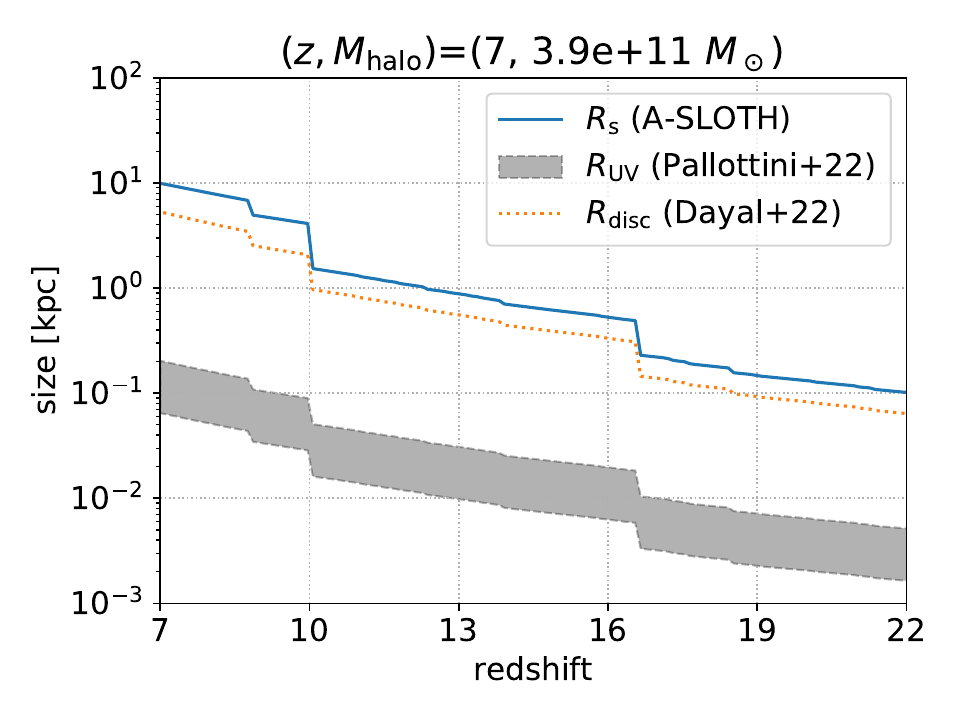}
    \caption{Comparison of the spatial extents of cold dust, for an example merger tree branch with target halo mass $3.9\times 10^{11}M_\odot$ at redshift $7$. We plot $R_{\rm s}$ assumed in \asloth as the extent of cold gas (solid), and the stellar half-light radius $R_{\rm UV}$ by \citet{Pallottini22} (equation \ref{eq:R_uv}; dashed), here for a broad range of stellar-to-halo mass ratio of $M_*=(10^{-4}$--$10^{-2})M_{\rm halo}$ (see also Figure \ref{fig:stellar_mass_metallicity}). For reference we also show the radii $R_{\rm disc}\equiv 0.18R_{\rm vir}$ employed as the dust extent in the semi-analytical model of \citet{dayal22}.}
    \label{fig:rUV_rdisk}
\end{figure}

The presence of dust can have a significant impact on the gas dynamics due to its large opacity to UV photons. Ejection of dust and gas can occur if the radiation pressure from the stars can unbind them from their gravitational potential \citep[e.g.,][]{Murray05,Thompson15}.

To take this into account, we newly incorporate in \asloth radiation-driven feedback onto dust and gas. We model the ejection in hot and cold phases separately, but keep the assumption that the dust and gas are dynamically coupled in each phase. We assume that the optical depth to photons is dominated by the dust
\begin{align}
    \tau_{\rm cold} &\approx \frac{3\kappa_{\rm d}M_{\rm d, cold}}{4\pi R_{\rm cd}^2} \sim 2.5 \left(\frac{\kappa_{\rm d}}{5\times 10^4{\rm cm^2\ g^{-1}}}\right)\left(\frac{M_{\rm d, cold}}{10^6\ M_\odot}\right) \left(\frac{R_{\rm cd}}{1\ {\rm kpc}}\right)^{-2}, \label{eq:tau_cold}\\
    \tau_{\rm hot} &\approx \frac{3\kappa_{\rm d}M_{\rm d, hot}}{4\pi R_{\rm vir}^2}, \label{eq:tau}
\end{align}
where $R_{\rm cd}$ is the extent of dust in cold ISM, and dust in hot ISM is assumed to co-exist with the gas in $R_{\rm vir}$. The opacity of dust is for simplicity set to a fixed value of $\kappa_{\rm d}\approx 5\times 10^4\ {\rm cm^2\ g^{-1}}$, inferred from the model by \cite{Draine84} at $10^{15}$ Hz where the (intrinsic) spectral energy distribution is expected to peak in high-redshift galaxies (Riaz et al. in prep.).

We consider momentum driven ejection \citep[e.g.,][]{Murray05} due to stellar radiation, and model the acceleration of dust and gas as
\begin{align}
    a_{\rm cold} \approx & \frac{{\rm min}[\tau_{\rm cold},1]L}{(M_{\rm cold}+M_{\rm d, cold})c}\\
    a_{\rm hot} \approx & \frac{{\rm min}\left[\tau_{\rm hot},1-{\rm min}[\tau_{\rm cold},1]\right]L}{(M_{\rm hot}+M_{\rm d, hot})c},
\end{align}
where $L$ is the bolometric luminosity from the stars present at time $t$. Here we assumed that the cold gas is embedded in the hot gas, and the hot gas can only receive momentum from photons that penetrate through the cold gas. We also neglected the momentum contributed from multiple scattering by dust-reprocessed infrared photons, as they are generally negligible \citep[e.g.,][]{Hopkins20}. We then define the timescales of radiation-driven ejection as
\begin{align}
    t_{\rm rad, cold} =& \frac{v_{\rm esc, cold}}{a_{\rm cold}}= \frac{\sqrt{2E_{\rm bind, cold}/M_{\rm cold}}}{a_{\rm cold}}\\
    t_{\rm rad, hot} =&  \frac{v_{\rm esc, hot}}{a_{\rm hot}} = \frac{\sqrt{2E_{\rm bind, hot}/M_{\rm hot}}}{a_{\rm hot}},
\end{align}
where the numerator is the typical escape velocity of hot and cold gas, obtained from their respective binding energies $E_{\rm bind, hot}, E_{\rm bind, cold}$ which are also used for modelling SN feedback \citep[Appendix B of][]{Chen22}. It is assumed that the gas carries most of the mass and binding energy, which is generally the case for typical dust to gas mass ratios.  

\begin{table}
    \centering
    \begin{tabular}{c|cc}
       Model & $t_{\rm acc,\odot}$ (Myr) & Cold dust extent $R_{\rm cd}$\\
       \hline
       Extend-30Myr & 30 & $R_{\rm s}$ (equation \ref{eq:R_s})\\ 
       Extend-3Myr & 3 & $R_{\rm s}$ (equation \ref{eq:R_s}) \\
       Comp-30Myr & 30 & $R_{\rm UV}$ (equation \ref{eq:R_uv}) \\
       Comp-3Myr & 3 & $R_{\rm UV}$ (equation \ref{eq:R_uv}) 
    \end{tabular}
    \caption{Summary of the models we employ for the growth timescale and extent of dust in the cold ISM.}
    \label{tab:models}
\end{table}

As seen in equation (\ref{eq:tau_cold}), the optical depth to UV photons is sensitive to the assumed spatial extent of the dust in the cold ISM, which may depend on each galaxy's history. One possibility is that dust co-exists with the cold ISM, as seen in the current Milky Way disc. This results in the spatial extent to be of a few 10 per cent of the virial radius, similar to that adopted in \citet{dayal22} when estimating the dust attenuation in $z\approx 7$ galaxies. Hints of disc-like structures are already seen in some $z>10$ galaxies by JWST \citep[e.g.,][]{Naidu22}, but it is difficult to draw a firm conclusion due to limited samples.

Another possibility is to assume that the dust is co-existent with the stars that produce them. The stars can occupy a much more compact radius than the cold ISM gas for high-z galaxies \citep[e.g.,][]{Yajima22}, which may be due to star formation occurring inside-out \citep{Baker23}. If stars and dust are dominantly produced from a single star-formation episode and have not experienced much feedback, the dust may follow the stellar population instead of the cold ISM.

With these possibilities in mind we consider two models for the spatial extent of dust in our work, which we name as the ``Extended" model and the ``Compact" model. The extended model assumes that the extent of dust is equivalent to that of the cold gas assumed in \asloth, i.e. $R_{\rm cd}=R_{\rm s}$. On the other hand, the compact model assumes that the dust extent is equivalent to the half-light radius of the galaxy. In this work we adopt a formula
\begin{eqnarray}
R_{\rm cd} = R_{\rm UV} \equiv 0.0745\ {\rm kpc}\left(\frac{M_*+10^{4}M_\odot}{10^8\ M_\odot}\right)^{0.249} \left(\frac{1+z}{8.7}\right)^{-1.1},
\label{eq:R_uv}
\end{eqnarray}
where we used the size-stellar mass relation at redshift $7.7$ obtained from zoom-in cosmological simulations by \cite{Pallottini22}, and the redshift evolution of the effective radius from observations over a wide range of $z=0$--$12$ \citep{Shibuya15, Ono23}. We added $10^4\ M_\odot$ in the stellar mass from the fit in \cite{Pallottini22}, which is roughly the resolution of stellar particles in their simulation.

 \begin{figure*}
   \centering
    \begin{tabular}{cc}
     \begin{minipage}[t]{0.5\hsize}
    \centering
    \includegraphics[width=\linewidth]{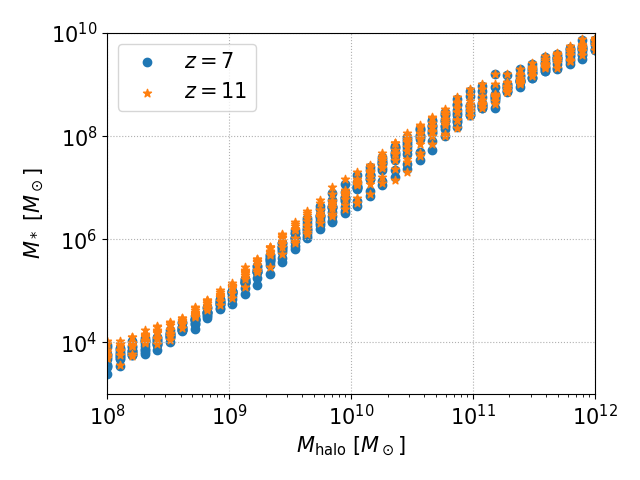}
    \end{minipage}
     \begin{minipage}[t]{0.5\hsize}
   \centering
    \includegraphics[width=\linewidth]{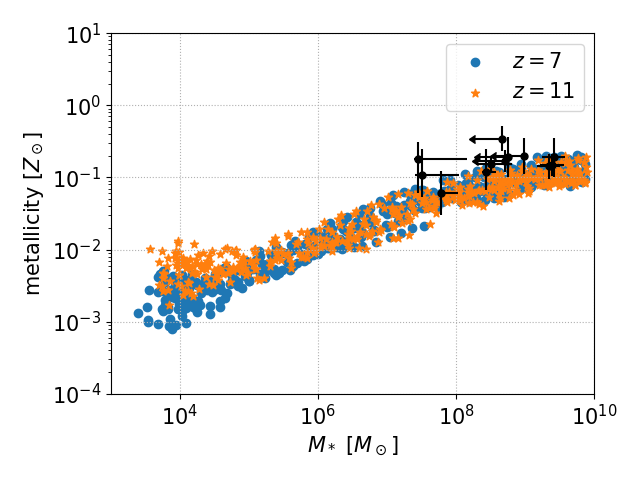} 
    \end{minipage}
    \end{tabular}
\caption{Summary of the stellar mass and gas-phase metallicity in the ISM of the sampled galaxies, for the Comp-30Myr model. The black points with errorbars in the right panel show the observed relation between stellar mass and metallicity for galaxies in the redshift range 6.5--7.5 \citep{Nakajima23}.}
 \label{fig:stellar_mass_metallicity}
 \end{figure*}
 
To illustrate the discrepancy of these two length scales, in Figure \ref{fig:rUV_rdisk} we plot them along a merger tree branch, for a galaxy with $M_{\rm halo, f}=3.9\times 10^{11}\ M_\odot$ at redshift 7. As the stellar masses depend on the employed models in \asloth, only for this plot we scale this by the halo mass and consider a wide range $M_*=(10^{-4}$--$10^{-2})M_{\rm halo}$ to see the rough values of $R_{\rm UV}$ (see also Figure \ref{fig:stellar_mass_metallicity}). The virial radii and $R_{\rm UV}$ have weak dependencies on the masses as $M_{\rm halo}^{1/3}$ and $M_*^{0.249}$ respectively, so the degree of discrepancy would only weakly depend on the adopted halo mass or stellar-to-halo ratio.

As shown in Figure \ref{fig:rUV_rdisk}, we find that the two radii differ by 1.5-2 orders of magnitude. This can translate to 3-4 orders of magnitude difference on the optical depth when a spherically symmetric dust distribution is assumed. We also draw the extent of dust $0.18R_{\rm vir}$ independently assumed in \cite{dayal22} for a semi-analytical modelling of dust extinction. This prescription is between the two cases considered in our work, but is much closer to $R_{\rm s}$.

In summary, our model parameters are summarized in Table \ref{tab:models}. We aim to find the parameter set(s) where we can reproduce both the dusty galaxies in ALMA and dust-free galaxies in JWST.

\section{Results}
\label{sec:results}

\subsection{General Properties of High-redshift Galaxies}
Before presenting the result of our dust modelling, we first summarize the general properties of the galaxy samples obtained by \asloth. We focus on the stellar masses and metallicities of the galaxies, which would be an important tracer of the production and growth of dust. 

Figure \ref{fig:stellar_mass_metallicity} shows the stellar masses and ISM metallicities of the galaxy samples for redshifts 7 and 11, for the Compact-30Myr model. The correlations are almost identical for the other models. The relation between the stellar and halo mass, shown on the left panel, depend weakly on the redshift. An exception is at the low mass-end of $M_{\rm halo}\lesssim 10^9\ M_\odot$, where the stellar mass is a factor of a few larger at $z=11$ than $z=7$. This difference is attributed to the relatively larger contribution from Pop-III star formation at higher redshifts \citep[e.g.,][]{riaz22}, that \rev{has higher efficiency than Pop-II star formation with $\eta_{\rm III}=2\eta_{\rm II}$}. Since the Pop-III IMF adopted in \asloth is top-heavy, this also leads to more SN and thus enhanced metallicity for a fixed stellar mass, as seen in the right panel. However, in practice these Pop-III dominated galaxies would be too dim to observe \citep{riaz22}, unless they may be strongly lensed with high magnification of $\sim 10^3$--$10^4$.

We compare the stellar-mass metallicity relation in the right panel to the measurements of \cite{Nakajima23}. As we adopt a solar metallicity of $Z_\odot=0.0134$ in \citet{Asplund09}, we use the corresponding oxygen abundance of $12+\log [{\rm O/H}]=8.69$ to convert the measured $12+\log [{\rm O/H}]$ to metallicity $Z$. While \asloth is not calibrated by the metallicity history, the simulated values reasonably agree with the observed samples at the higher end of $M_*$. 

\subsection{Evolution of Dust Mass}
Here we look into the redshift evolution of the dust masses for galaxies of two representative halo masses, and assess the relative importance of the processes included in our model.

 \begin{figure*}
   \centering
    \begin{tabular}{cc}
     \begin{minipage}[t]{0.5\hsize}
    \centering
 \includegraphics[width=\linewidth]{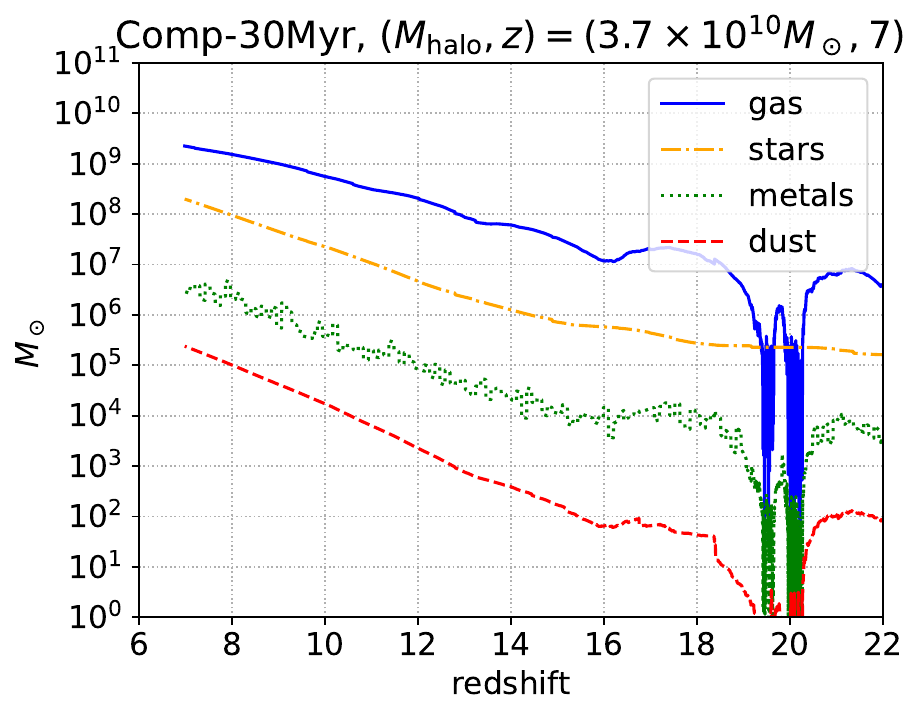}
    \end{minipage}
     \begin{minipage}[t]{0.5\hsize}
    \centering
 \includegraphics[width=\linewidth]{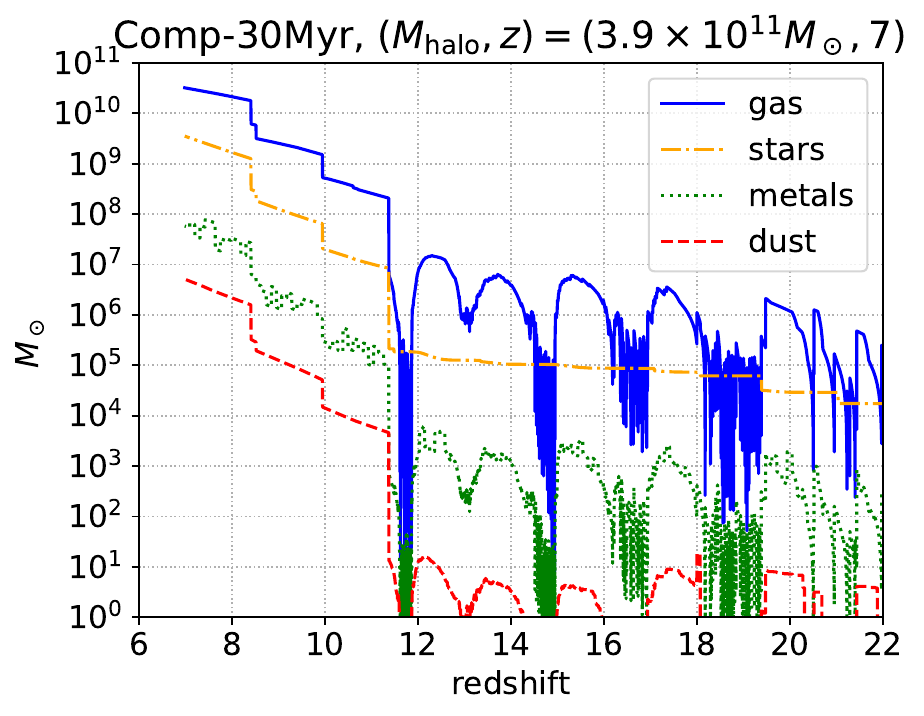}
    \end{minipage}
    \end{tabular}
\caption{Redshift evolution of the masses of each component for the Compact-30Myr model, for two merger tree realizations with halo masses of $3.7\times 10^{10}\ M_\odot$ and $3.9\times 10^{11}\ M_\odot$ at redshift $7$. The oscillation of gas and metal masses at the early phases of $z\gtrsim 15$ reflect the stochastic processes of SN feedback. We note that the plot follows a single representative merger tree branch defined by \asloth, and the jumps in the masses are due to merger with another halo from another branch.}
\label{fig:z_gas_dust}
 \end{figure*}

Figure \ref{fig:z_gas_dust} shows the evolution of the masses of dust as well as other baryonic components along a single merger tree branch, with final halo masses at redshift 7 of $3.7\times 10^{10}M_\odot$ and $3.9\times 10^{11}M_\odot$. The gas mass, corresponding to the sum of the masses of hot and cold gas, initially oscillates at high redshifts of $z\gtrsim 15$. This reflects the importance of SN feedback when the halo mass is still small and its gravitational potential is shallow. Since we resolve individual stars at each timestep, the feedback is stochastic and creates oscillations in the gas mass as well as those of metals and dust (see also Figure 3 of \citealt{Hartwig22}). Later on, the mass of each component grows smoothly, but sometimes increases discontinuously due to mergers with other haloes. As the redshift decreases, the stellar mass grows slightly faster than the gas mass. The dust mass generally grows in accord to the stellar mass, reflecting the SN yield. A qualitatively similar evolution of these masses is found irrespective of the halo mass, as shown for two cases of halo mass shown in this figure.

\begin{figure*}
    \centering
    \begin{tabular}{cc}
     \begin{minipage}[t]{0.5\hsize}
    \centering
 \includegraphics[width=\linewidth]{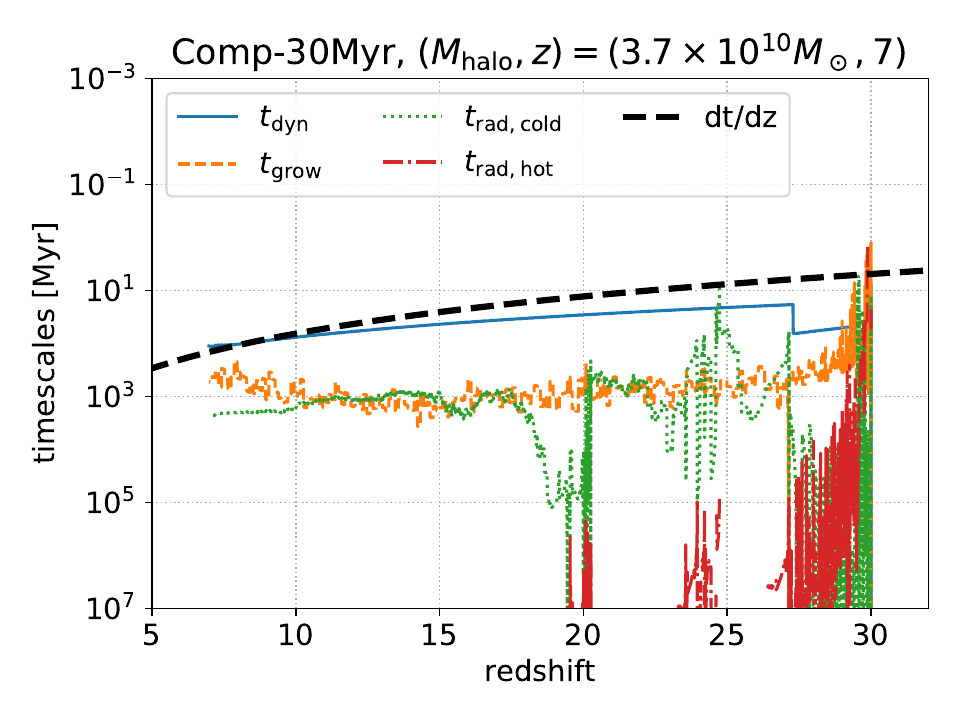}
    \end{minipage}
     \begin{minipage}[t]{0.5\hsize}
    \centering
 \includegraphics[width=\linewidth]{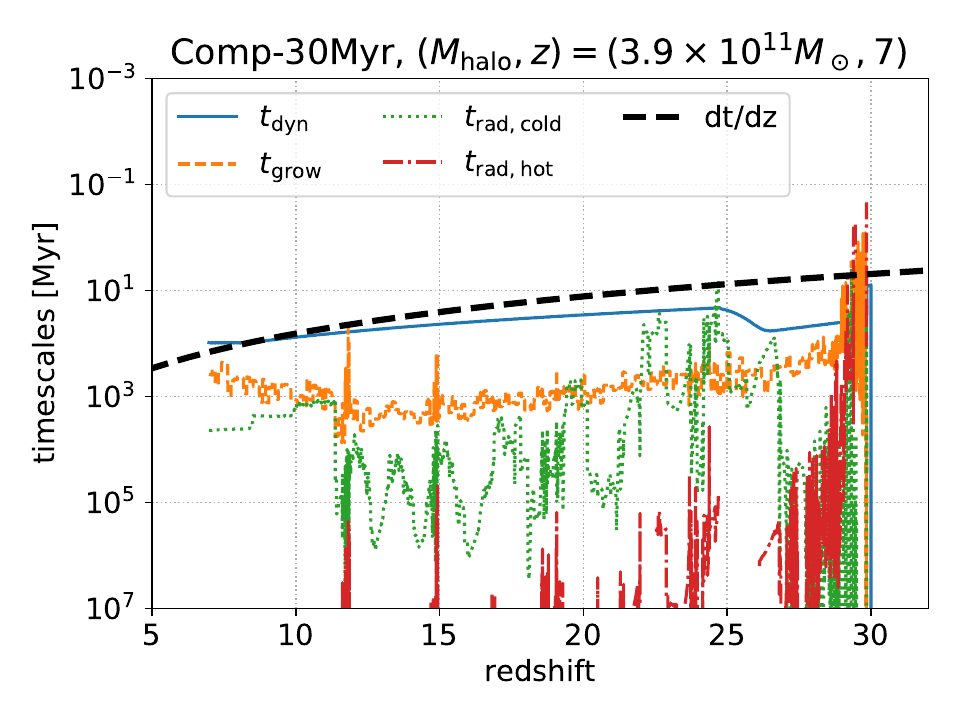}
    \end{minipage}
    \end{tabular}
    \caption{Timescales of dust growth and ejection processes compared to the dynamical time (solid line) and cosmic time (thick dashed line), for the Compact-30yr model. Note the inverted y-axis, which makes the process for the line on top most important. The two merger tree realizations are the same as those in Figure \ref{fig:z_gas_dust}.}
    \label{fig:timescales}
\end{figure*}

We next plot in Figure \ref{fig:timescales} the various timescales regarding dust outlined in Section \ref{sec:Methods}, as well as the dynamical timescale $t_{\rm dyn}$ and elapsed time per unit redshift $dt/dz$. Note that the line on the top denotes the most important process that governs dust evolution. We adopt the same two merger tree branches as in Figure \ref{fig:z_gas_dust}. We find that the dust growth can become important only for high-mass haloes at late times, when the metallicities reach $\sim 0.1Z_\odot$.

The radiative feedback on dust is found to be generally unimportant, except for the very beginning of galaxy formation at $z\approx 30$. The timescales $t_{\rm rad, cold}$ occasionally has spikes due to enhanced star formation or gas clearing by SNe, but is always a few orders of magnitude longer than both the dust growth timescale and the dynamical time. This validates the assumption in the previous semi-analytical models neglecting radiative feedback, which reproduces the dust properties of rather normal star-forming galaxies at $z<10$. However the radiative pressure can still be important in the case of bursty star formation, with brief periods of high star-formation efficiency. We demonstrate this in detail in Section \ref{sec:jwst}, where we apply our model to the earliest galaxies at $z>10$ found by JWST.

\subsection{Comparison with ALMA galaxies at redshifts $\approx 7$}
\label{sec:alma}

\begin{figure}
    \centering
    \includegraphics[width=\linewidth]{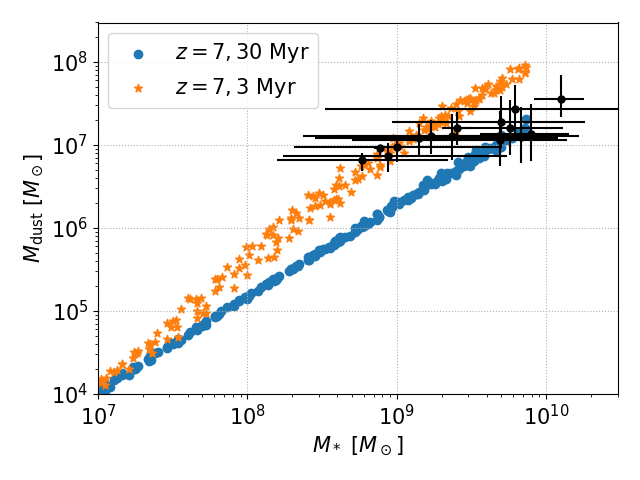}
    \caption{The dust masses of the galaxies at redshift 7, for the Compact dust model with different growth timescales $t_{\rm acc,\odot}=30$ Myr and $3$ Myr. The black points with error bars on the top right shows the observed samples by ALMA in the redshift range 6.5-7.5 \citep{Knudsen17,Hashimoto19,Sommovigo22}.}
    \label{fig:dust_z7}
\end{figure}

Figure \ref{fig:dust_z7} shows the dust masses obtained by our model at redshift 7. This is compared to the observed galaxy samples in the redshift range 6.5--7.5 \citep{Knudsen17,Hashimoto19,Sommovigo22}, shown as black points with error bars. \rev{For sources where dust emission is observed in a single frequency band, the measured dust masses can have significant uncertainties since it sensitively depends on the assumed dust temperature \citep[e.g.,][]{Sommovigo20,Cochrane22}.} Furthermore, representing the dust by a single temperature may not always be a good approximation, as there is no reason to expect for multiple ISM components to have similar dust temperatures. The dust in the cold ISM, where it can be irradiated by stars, may have higher temperatures than those in the diffuse ISM. This may partially explain the apparently larger scatter seen in the observed points than ours. While these caveats should be taken into consideration in future works, our estimates are roughly in agreement with the dust masses inferred from ALMA observations. \rev{Our model with two cases of $t_{\rm acc, \odot}=3$ and $30$ Myr reproduces the range in dust-to-stellar mass ratio of $M_{\rm dust}/M_*\approx 10^{-3}$--$10^{-2}$ in these samples, as well as the (crude) upper limits on $M_{\rm dust}$ for galaxies with non-detection of dust continuum (Appendix \ref{sec:alma_upper_limits}).}

Comparing the $t_{\rm acc, \odot}=3$ Myr model to the 30 Myr model for a fixed $M_*$, the \rev{two dust masses} overlap at \rev{low stellar masses} of $M_*\lesssim 10^7M_\odot$, \rev{while they start} to deviate at higher masses by up to a \rev{factor of $5$}. This is due to the correlation of the metallicity with the stellar mass seen in Figure \ref{fig:stellar_mass_metallicity}. As the cosmic age at $z=7$ is $\approx 800$ Myr, the difference in $t_{\rm acc, \odot}$ will \rev{greatly affect the dust mass} for large galaxies at $Z\sim 0.1Z_\odot$ while it will be irrelevant for low-mass galaxies of $Z\sim 0.01Z_\odot$ (see also Figure \ref{fig:timescales}). 

\rev{We also investigate the relation of $M_*$ and $M_{\rm dust}$ for the Extended dust models, and find that it is almost identical to the Compact dust model in Figure \ref{fig:dust_z7}. This is because the radiative feedback in the model is weak even for the Compact model (see the timescales in Figure \ref{fig:timescales}), that it would barely influence the evolution of the dust mass. This results in negligible difference in the final dust mass between the Compact and Extended models.}

\section{Comparison with JWST galaxies at redshifts >10: Requirement of Efficient Star-formation}
\label{sec:jwst}
At the time of writing, there are no confident detections of dust emission from JWST galaxies at $z>10$. Observations of dust continuum emission by ALMA have been reported for two galaxies GHZ1/GLASS-z11 at $z=11.0$ \citep{Yoon22} and GHZ2/GLASS-z12 at $z=12.1$ \citep{Bakx23}, with upper limits on dust mass of $\lesssim 10^4\ M_\odot$ and $<1.5\times 10^6\ M_\odot$ respectively. The former especially indicate a severe deficiency of dust in these high-z galaxies, with dust-to-stellar mass ratio of $M_{\rm dust}/M_* \lesssim 10^{-5}$--$10^{-4}$. This is much smaller than the ratio $10^{-3}$--$10^{-2}$ seen in the local universe as well as the $z\approx 7$ samples in Section \ref{sec:alma}.

In addition to the ALMA non-detections, lack of dust is independently implied from photometric observations by the blue slope in the rest-frame UV part of the SED \citep[e.g.,][]{Finkelstein22,Atek23,Furtak23}. This has been translated to measurements or upper limits on the dust attenuation, for a handful of JWST galaxies. Here we compute the V-band attenuation $A_V$ from dust following \cite{Ziparo23}, using the relation between $A_V$ and the dust optical depth at 1500\AA~for dust composition similar to the Milky Way \citep{Ferrara22b}
\begin{align}
    A_V \approx  0.4\tau_{\rm 1500} \sim 2.5\left(\frac{M_{\rm d, cold}}{10^6\ M_\odot}\right) \left(\frac{R_{\rm cd}}{1\ {\rm kpc}}\right)^{-2},
    \label{eq:A_V}
\end{align} 
where we multiplied a factor of 3 from the equation of \cite{Ziparo23} to apply to a uniform spherical dust distribution assumed here. We compare our model galaxies at $z=11$ to six spectroscopically confirmed high-z galaxies with available dust attenuation measurements, as well as candidate galaxies with dust mass constraints from ALMA follow-up, and with high dust attenuation (Table \ref{tab:Mstar_dust_jwst}).

\begin{table*}
    \centering
    \renewcommand{\arraystretch}{1.5}
    \begin{tabular}{c|ccccc}
       Spectroscopically Confirmed & $z$ & log$_{10}[M_*/M_\odot]$ & $\dot{M}_*$ [$M_\odot$ yr$^{-1}$] & $A_V$ & $M_{\rm dust}$ [$M_\odot$]\\
       \hline
       GN-z11$^{*\rm a, \dag}$ & 10.6 & $8.73^{+0.06}_{-0.06}$&  $18.78^{+0.81}_{-0.69}$  & $0.17^{+0.03}_{-0.03}$& -- \\
       Maisie's Galaxy$^{*\rm b}$ & 11.44 & $8.4^{+0.3}_{-0.4}$ & $3^{+2}_{-2}$  & $0.07^{+0.09}_{-0.05}$ & -- \\
       GS-z10$^{*\rm c}$ & 10.38 & $7.9^{+0.3}_{-0.5}$&  $1.0^{+0.6}_{-0.5}$  & $0.2^{+0.3}_{-0.1}$& -- \\
       GS-z11$^{*\rm c}$ & 11.58 & $8.9^{+0.2}_{-0.4}$ &  $2.0^{+3.0}_{-1.8}$    & $0.2^{+0.3}_{-0.2}$& -- \\
       GS-z12$^{*\rm c}$ & 12.63 & $8.4^{+0.3}_{-0.7}$&  $1.3^{+1.9}_{-0.9}$   & $0.2^{+0.4}_{-0.2}$& -- \\
       GS-z13$^{*\rm c}$ & 13.20 & $7.7^{+0.4}_{-0.4}$ &  $1.0^{+1.0}_{-0.5}$    & $0.3^{+0.4}_{-0.2}$& -- \\ \hline
       Candidates & & & &  \\ \hline 
       GHZ1/GL-z11$^{*\rm d,e,f}$ & 10.4$^{+0.4\ *{\rm d}}_{-0.5}$, 10.47$^{+0.38\ *{\rm e}}_{-0.89}$ & $9.6^{+0.2\ *{\rm d}}_{-0.4}$, $9.1^{+0.2\ *{\rm e}}_{-1.0}$  & $10^{+17\ *{\rm d}}_{-5}$, $10.7^{+42.7\ *{\rm e}}_{-4.7}$   & $0.3^{+0.4{\ *{\rm d}}}_{-0.2}$ & $\lesssim 10^{4\ *{\rm f}}$\\
       GHZ2/GL-z12$^{*\rm d,g,h}$ & 12.4$^{+0.1\ *{\rm d}}_{-0.3}$, $12.11^{*{\rm g \ (fixed)}}$ & $9.1^{+0.3\ *{\rm d}}_{-0.4}$, $8.1^{+0.4\ *{\rm g}}_{-0.1}$ &  $6^{+5\ *{\rm d}}_{-2}$, $19.1^{+13.5\ *{\rm g}}_{-10.1}$  & $0.1^{+0.2{\ *{\rm d}}}_{-0.1}$ & $<1.5\times 10^{6 \ *{\rm h}}$ \\
       PENNAR$^{*\rm i}$ & 12.11$^{+0.12}_{-0.15}$ & 9.11$^{+0.34}_{-0.16}$ & $13^{+19}_{-4.1}$  &    2.36$^{+0.14}_{-0.12}$ & --
    \end{tabular}
	\caption{The $z>10$ galaxies and candidates observed by JWST with dust mass/attenuation measurements, as well as a candidate with high dust attenuation. The references to the values quoted in this table are below: a \citep{Bunker23}; b \citep{Haro23}; c \citep{Robertson23}; d \citep{Naidu22}; e \citep{Castellano23}; f \citep{Yoon22}; g \citep{Santini23}; h \citep{Bakx23}; i \citep{Rodighiero23}. We omitted the ``KABERLABA" candidate in \citet{Rodighiero23}, as we do not simulate galaxies with stellar masses measured for this candidate ($\log_{10}M_*=11.29^{+0.38}_{-0.51}$). $\dag$ For GN-z11, the values in the table assume that stellar radiation is responsible for the observed light. There is evidence for an active galactic nuclei powering UV emission \citep{Bunker23,Scholtz23}, which can significantly change the estimates of $M_*$ and $\dot{M}_*$.}
    \label{tab:Mstar_dust_jwst}
\end{table*}

\begin{figure*}
    \centering
    \includegraphics[width=\linewidth]{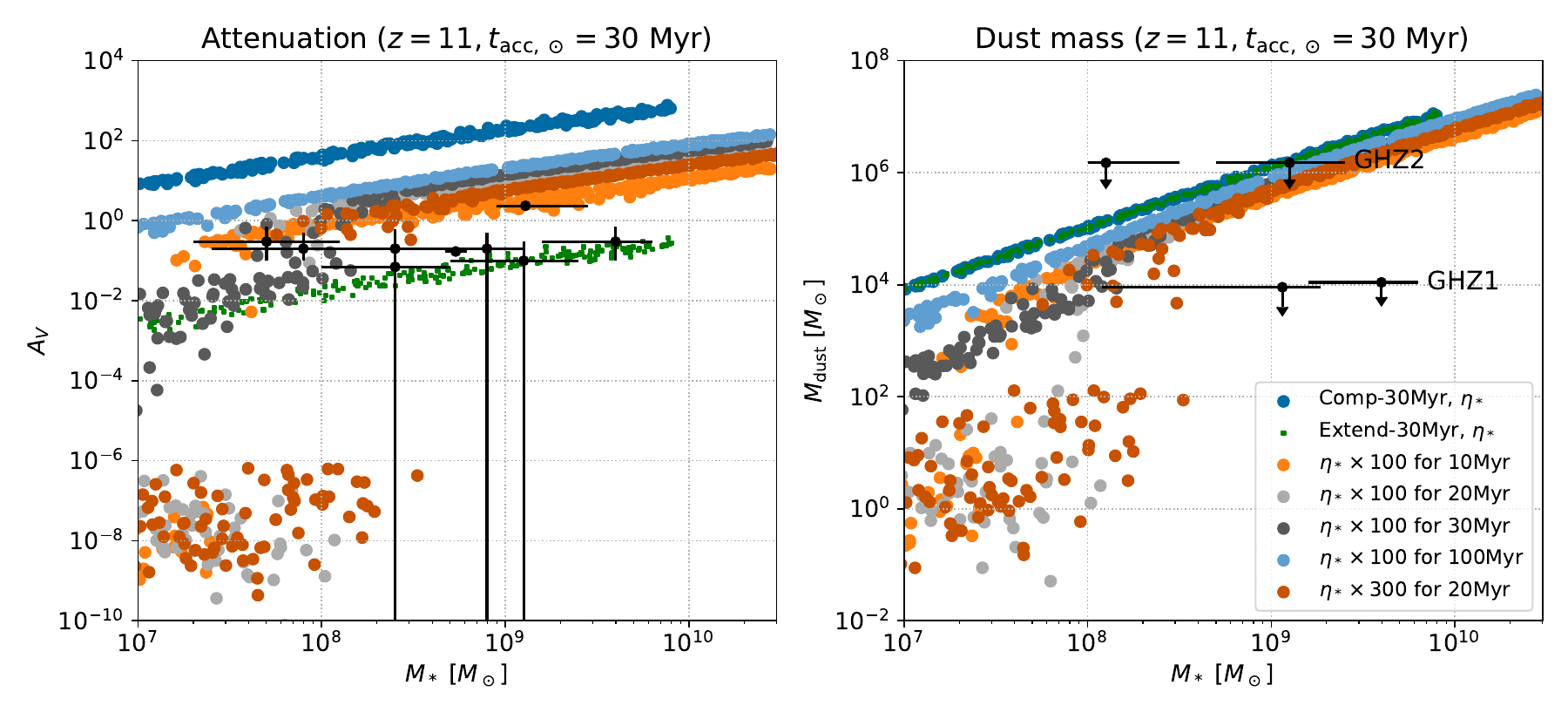}
    \caption{The attenuations and dust masses of the galaxies at redshift 11 as a function of $M_*$, focusing on the model with $t_{\rm acc,\odot}=30$ Myr. \rev{The small squares are for the Extended dust model, and the other large circles are for the Compact dust model.} In the right panel we show two independent measurements of $M_*$ carried out for both GHZ1 and GHZ2, where the more recent analyses resulted in lower mass estimates. For GHZ1, we intentionally displace the upper limit of $M_{\rm dust}$ by $10\%$ for visibility of the two error bars on $M_*$.}
    \label{fig:dust_z11}
\end{figure*}

We present the dust masses and the attenuation in Figure \ref{fig:dust_z11}. As seen from the figure, our model \rev{for the Compact dust distribution} results in much higher attenuation and dust masses than the JWST galaxies with low dust attenuation. \rev{One may be able to reduce the value of $A_V$ in agreement with observations by modifying the spatial distribution of dust, either adopting the Extended dust model with larger $R_{\rm cd}$ (smaller $\tau_{\rm cold}$; see left panel), or considering asymmetric/clumpy structure as suggested by recent rest-frame far infrared observations of high-redshift galaxies at $z=6$--$8$ \citep[e.g.,][]{Harikane20,Ferrara22b, Tamura23}. However, both modifications cannot reduce the dust masses, which is yet orders of magnitude higher than the upper limit for GHZ1.}

This discrepancy in dust mass stems from the fact that most of the dust cannot be ejected by radiation pressure, as hinted in Figure \ref{fig:timescales}, even for the compact dust model that predicts higher optical depth. This inefficient dust clearing can be understood by the following timescale argument. The radiation pressure is roughly proportional to the star-formation rate $\dot{M}_*$, and using the formulations in \citet{Hartwig22} this can be expressed as 
\begin{equation}
    \frac{L}{c}\approx \epsilon_{\rm II}\dot{M}_*c \approx \epsilon_{\rm II}\left(\eta_{\rm II}\frac{M_{\rm cold}}{\sqrt{4\pi/3}t_{\rm dyn}}\right)c,
\end{equation}
where $\eta_{\rm II}$ is the star formation efficiency of Pop II stars calibrated as $0.19$ \citep{Hartwig22}, and $\epsilon_{\rm II}\equiv L/(\dot{M}_* c^2)\sim 10^{-3}$ is the radiation conversion fraction for Pop II stars \citep{Murray05}, assuming a Kroupa IMF \footnote{We neglect the contribution from Pop III stars, as they are generally negligible at the redshift and stellar mass of interest \citep[see also][]{riaz22}.}. For $\tau_{\rm cold}>1$, which is the most optimistic case for dust clearing, we obtain the dust ejection timescale as
\begin{equation}
    t_{\rm rad, cold} \approx \frac{M_{\rm cold}v_{\rm esc, cold}}{L/c} \sim \frac{\sqrt{4\pi/3}}{\eta_{\rm II}}\frac{v_{\rm esc, cold}}{\epsilon_{\rm II} c} t_{\rm dyn}.
     \label{eq:trad_cold}
\end{equation}
The escape velocity is
\begin{align}
    v_{\rm esc, cold}\approx&\sqrt{\frac{2GM_{\rm halo}}{R_{\rm s}}} \nonumber \\
    \sim& 300\ {\rm km\ s^{-1}}\left(\frac{M_{\rm halo}}{10^{10}\ M_\odot}\right)^{1/2} \left(\frac{R_{\rm s}}{1\ {\rm kpc}}\right)^{-1/2},
\end{align}
so the factor $v_{\rm esc, cold}/\epsilon_{\rm II} c$ is expected to be of order unity. Since $t_{\rm dyn}$ is the timescale that cold gas is supplied from the reservoir of hot gas (equation \ref{eq:coldgas}), it would be difficult to clear out the cold gas (along with the dust) to masses much less than the hot gas, unless $\eta_{\rm II}\gg 1$. This is qualitatively consistent with the findings of \cite{Ziparo23} that bursty star formation (typically 10s of times of that derived from the Kennicutt-Schmidt relation) is necessary to eject the gas and dust.  
A bursty star formation history is indeed found to be common for high-redshift galaxies both from cosmological simulations and JWST observations \citep[e.g.,][]{Pallottini22,Yajima22,Dressler23,Looser23}, which can be due to the star-forming gas being highly clumpy, possibly in combination with lack of substantial feedback from SNe and stellar winds \citep{Dekel23,Renzini23}.

To investigate the effect of bursty star-formation on the dust, we conduct calculations mimicking starburst galaxies, with enhanced star formation efficiencies for a brief amount of time before the end of the calculation. Specifically, we raise both efficiencies $\eta_{\rm II}, \eta_{\rm III}$ by a factor of 100 and 300, \rev{for a certain period of time from 10 to 100 Myr before $z = 11$.} \rev{The choices of the enhancement factor are inspired from the cosmological simulations of \cite{Yajima22} that studied galaxies with masses and redshifts similar to our interest, where bursty star formation is seen along with a continuous one, with the former having star-formation rates of $\sim 100$ times the latter. They are also motivated by inspection of Figure \ref{fig:z_gas_dust}, where $t_{\rm rad, cold}$ is 1-2 orders of magnitude longer than $t_{\rm dyn}$ for the default values of $\eta_{\rm II}$ and $\eta_{\rm III}$.} The stellar age \rev{of the bursty component} is chosen to be consistent with measurements of JWST galaxies from SED fitting, although they can be quite sensitive to the adopted star-formation history.

The resulting modifications to the $M_*$--$A_V$ and $M_*$--$M_{\rm dust}$ relations are shown in Figure \ref{fig:dust_z11}. We find that for ``starbursts" of duration $<20$ Myr, dust is cleared for galaxies below a certain $M_*$, which is larger for longer duration and larger efficiencies. This directly reflects the scaling of equation (\ref{eq:trad_cold}), with the condition for dust to be cleared as $t_{\rm rad, cold}$ being less than the starburst duration. The much smaller change for massive galaxies of $M_*\gtrsim 10^9M_\odot$ is also explained from this equation by the longer $t_{\rm dyn}$ for more massive galaxies in general, as well as by the larger escape velocity of the dust and gas.

Once the duration exceeds $\gtrsim 30$ Myr, the dust masses return close to the original correlations without the enhancement of $\eta_{\rm II}, \eta_{\rm III}$. This is due to new dust being produced from SNe of stars formed in this starburst episode. This threshold timescale coincides with the lifetime of stars with lowest mass exploding as SNe ($10M_\odot$ in \asloth), which are most abundant for a Pop-II IMF. This indicates that radiative feedback is responsible for the dust clearing, rather than the SN feedback that produces new dust.

For these models, we find that a majority of the stellar mass is produced in this ``starburst" episode. The time-averaged star-formation \rev{rate} for a galaxy with stellar mass of a few $\times 10^8\ M_\odot$ and stellar age of a few $10$ Myr is $\approx 10\ M_\odot {\rm yr}^{-1}$. This is on the high end but consistent with the star-formation rates of the observed JWST galaxies, most notably with GHZ1 with a remarkably low dust mass. 

A detailed comparison of dust mass with GHZ1 is not straightforward due to the uncertain (and disparate) stellar mass measurements. Adopting the more recent measurement of \cite{Castellano23} that took into account lensing magnification, the requirement for the enhancement in the efficiency is at least $\sim 100$ times of the original value calibrated for normal (lower redshift) galaxies. If we attribute this to $\gtrsim 100$ times shorter free-fall timescale of the star-forming gas, this indicates a $10^4$--$10^5$ higher density, likely in the form of clumps. The resulting gas density is roughly consistent with the required density for a feedback-free star formation, that is proposed to explain the luminous JWST galaxies \citep{Dekel23}.

\rev{The link between bursty star-formation and dust clearing has also been suggested by recent JWST observations. \cite{Topping23} analyzed the rest-UV slopes $\beta_{\rm UV}$ of 179 $z>9$ galaxies, finding that the subset of 44 galaxies with extremely blue colours ($\beta_{\rm UV}<-2.8$) have a higher specific star-formation rate (median 79 Gyr$^{-1}$) and lower stellar mass (median $\log (M_*/M_\odot)=7.5$) than the entire sample. Analysis of JWST data on the redshift $9.11$ galaxy MACS1149-JD1 also found a dust-free environment, with a major starburst episode $\sim 10$ Myr ago and a relatively low stellar mass of $1.6^{+0.3}_{-0.3}\times 10^8 M_\odot$ \citep{Stiavelli23}. These results are in line with our findings that low stellar mass and short burst duration are both favorable for dust ejection.}

\section{Discussion and Conclusion}
\label{sec:conclusion}
In this work we presented our \rev{update} to the \asloth semi-analytical model \citep{Hartwig22}, that newly tracks dust evolution in the early universe. We have considered several models governing the growth and ejection of dust, to compare our model against the observations of high-redshift galaxies by ALMA and JWST.

Our models of dust mass show good agreement when compared to galaxies at redshift $z=7$ observed by ALMA. However, when looking at even earlier galaxies with $z>10$ detected by JWST, we find that bursty star formation, with stellar ages $<30$ Myr, is likely required to explain the absence of dust. Our analysis suggests that the timing and strength of bursty star formation can greatly change the amount of dust. Specifically, our analysis shows that the low dust attenuation and weak \rev{rest-frame far-infrared} dust emission in JWST samples ($z > 10$) may be indicative of brief starbursts with efficiencies much larger than the calibrated values in \asloth. 

\subsection{Prospects and Implications for Future Observations}
\begin{figure*}
    \centering
    \includegraphics[width=\linewidth]{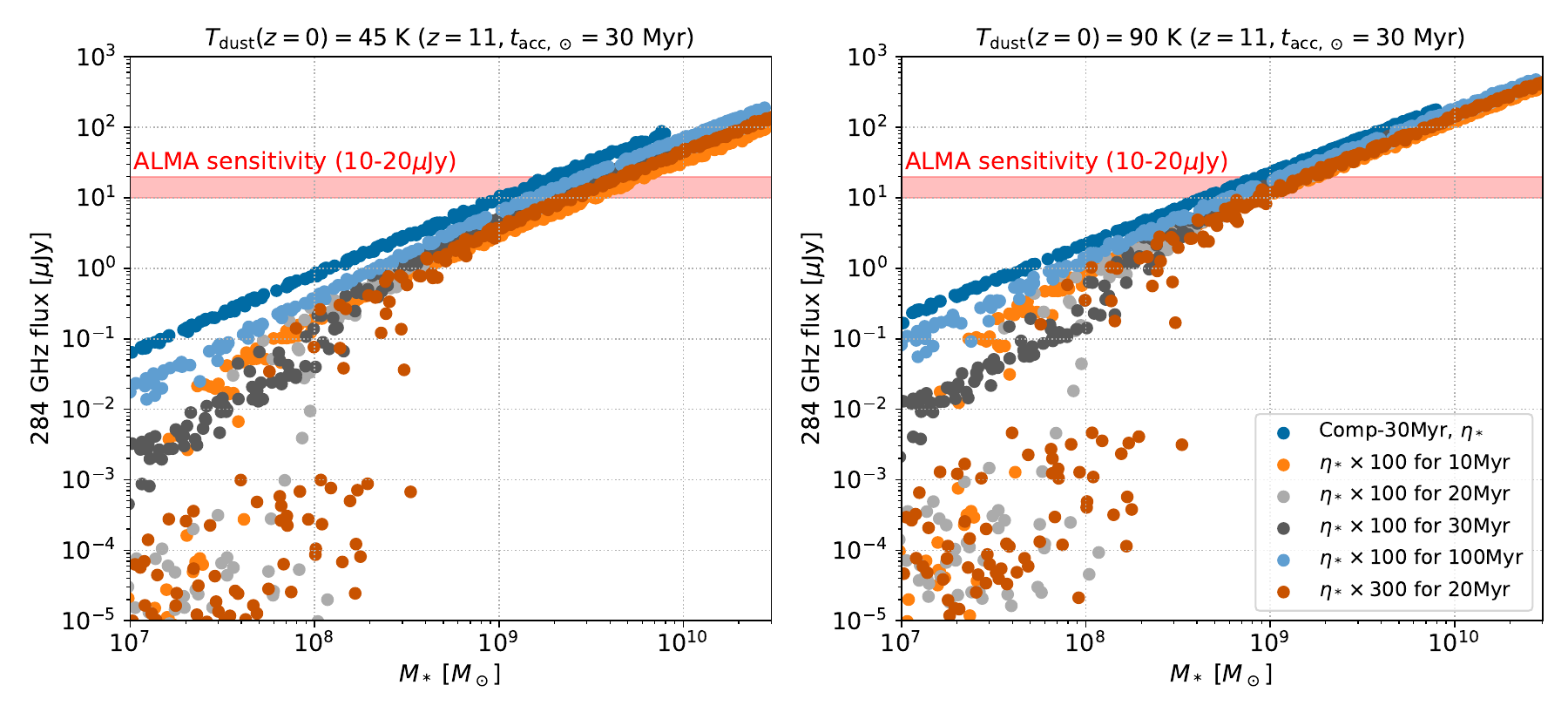}
    \caption{\revrev{The expected flux of dust emission at $284$ GHz (rest frame 88$\mu$m)} at redshift 11, for two representative dust temperatures $45$ K and $90$ K (see also equation \ref{eq:T_dust_z} and main text). \revrev{The models are the same as} those in Figure \ref{fig:dust_z11}, except that we removed the Extend-30Myr model as it gives a prediction almost identical to the Comp-30Myr model. The red shaded region shows the \revrev{typical ALMA sensitivity ($10$--$20\ \mu$Jy) from the past follow-ups} of high-redshift candidates \citep{Yoon22,Bakx23}.}
    \label{fig:ALMA_z11}
\end{figure*}
\rev{Our model implies that while galaxies with low $M_*$ are more susceptible to dust ejection by bursty star formation, mature galaxies with high $M_*$ are more likely to retain a significant amount of dust. Such dusty, mature galaxies at $z>10$ may be detectable in rest-frame far infrared, like the dusty galaxies at $z\approx 7$. Here we estimate the detectability of dust emission from such galaxies by ALMA, focusing on the case of $z=11$.}

\rev{For a galaxy of dust mass $M_{\rm dust}$ and a single representative dust temperature $T_{\rm dust}$, the observed flux at frequency $\nu=\nu_{\rm obs}$ is given as \citep[][Appendix A]{Yoon22}
\begin{equation}
    S_{\rm obs} = (1+z)\kappa_{\nu_{\rm rest}} 4\pi B_{\nu_{\rm rest}}(T_{\rm dust})\frac{M_{\rm dust}}{4\pi D_L^2},
    \label{eq:S_obs}
\end{equation}
where $\nu_{\rm rest}=\nu_{\rm obs}(1+z)$, $B_{\nu}(T)$ is the Planck function, $\kappa_\nu$ is the opacity, and $D_L\approx 120$ Gpc is the luminosity distance at $z=11$. We adopt $\nu_{\rm rest}=3410$ GHz ($\nu_{\rm obs}=284$ GHz) corresponding to the frequency of the \OIII 88$\mu$m line, that has been simultaneously probed with the dust continuum in past ALMA follow-up of GHZ1 and GHZ2 \citep{Yoon22,Bakx23}. The opacity $\kappa_\nu$ is expressed as $\kappa_{\nu}=\kappa_{\nu_0}(\nu/\nu_0)^\beta$, where we adopt $\kappa_{\rm \nu_0}=0.484$ cm$^2$ g$^{-1}$ at $\nu_0=352.697$ GHz ($850\ \mu$m), and $\beta=1.6$ following \cite{Yoon22}. The frequency-integrated infrared luminosity is then $L_{\rm IR}\approx 5.98\times 10^{-6}L_\odot (M_{\rm dust}/M_\odot)(T_{\rm dust}/{\rm K})^{5.6}$.}

\rev{We include heating by the cosmic microwave background (CMB), which raises $T_{\rm dust}$ from $T_{\rm dust}(z=0)$, the dust temperature had the galaxy been at $z=0$ \citep{daCunha13}. We adopt $T_{\rm dust}(z=0)$ of $45$ and $90$ K, values inferred from the galaxies at $z\approx 7$ \citep{Sommovigo22} and GHZ1 \citep{Yoon22} respectively. We furthermore cap the ($z=0$) dust temperature so that $L_{\rm IR}$ does not exceed the luminosity $L$ produced by the stars\footnote{\rev{In reality, $T_{\rm dust}$ can be self-consistently determined from energy balance between UV/optical absorption and IR re-emission, by solving radiative transfer assuming a dust geometry and an attenuation law. Here, for simplicity these uncertainties are absorbed in the parameter $T_{\rm dust}(z=0)$.}}. In summary, $T_{\rm dust}$ in equation (\ref{eq:S_obs}) is obtained by
\begin{eqnarray}
    T_{\rm dust}^{5.6}(z)&=&\left\{{\rm min}\left[T_{\rm dust}(z=0), \left(\frac{L/5.98\times 10^{-6}L_\odot}{M_{\rm dust}/M_\odot}\right)^{1/5.6}{\rm K}\right]\right\}^{5.6} \nonumber \\
    &&- T_{\rm CMB}^{5.6}(z=0) + [(1+z)T_{\rm CMB}(z=0)]^{5.6}  ,
    \label{eq:T_dust_z}
\end{eqnarray}
where $T_{\rm CMB}(z=0)=2.725$ K is the CMB temperature at $z=0$.}

\rev{Figure \ref{fig:ALMA_z11} shows the expected dust flux at $z=11$, for different masses as a function of $M_*$. For galaxies with $M_*\gtrsim 10^9M_\odot$, the dust emission can be detectable assuming a typical sensitivity of $10$--$20\ {\rm \mu}$Jy in previous follow-ups \citep{Yoon22,Bakx23}. While our estimates are crude and dependent on the dust temperature that is unconstrained in this work, they motivate future ALMA follow-ups of high-redshift mature galaxies to probe the dust evolution in the early universe.}

\rev{Our results can also be important for interpretations of high-redshift galaxies that require accurate modeling of dust attenuation/emission. One example is emission line diagnostics, where upcoming multi-wavelength observations from JWST and ALMA will provide an unprecedented opportunity to study the physical properties of high-redshift galaxies in detail.}
However, accurate interpretation of the data, such as inferring line ratios and estimating metallicities, will require careful consideration of the effects of dust attenuation. 
The use of multi-line ratio diagnostics, such as the \OIII 88\mum/5007\AA\, ratio, has been proposed theoretically as a way to probe the metallicity of high-redshift galaxies \citep{Nakazato23}, but \rev{the presence of dust studied in this work and its unknown geometry can limit the accuracy of such measurements}. Similarly, estimates of electron density based on the \OIII 88\mum / 5007 \AA\, ratio \citep{Fujimoto22} can be affected by dust attenuation, leading to potential biases in the inferred physical properties of high-redshift galaxies. Furthermore, it should be noted that the dust attenuation models calibrated toward observed galaxies in the epoch of reionization \citep[e.g.][]{Shimizu14, Sun16} may not be applicable to the galaxies even further $z \gtrsim 10$. Our study highlights the need for further refinement of the methods for estimating dust \rev{attenuation and emission} in high-redshift galaxies, and contribute to a better understanding of galaxy formation and evolution in the universe 500 Myr after the Big Bang.

\subsection{Possible Caveats}
We conclude by listing the possible caveats of our work that can be improved in future studies. We have neglected the contribution from asymptotic giant branch (AGB) stars \citep{Ferrarotti06,Zhukovska08,Ventura12,Nanni13,Dellagli17} in the dust yield. This would be important at low redshifts, but we expect that this would not greatly affect our results as we focus on high-redshift galaxies at $z\geq 7$ \citep[e.g.,][]{Mancini15}. An inclusion of AGB yields would be important if we wish to calibrate the dust properties at low redshift, like in the Milky Way \citep[e.g.,][]{deBennassuti14}. While one advantage of \asloth is following the lives of individual stars, it would be computationally more expensive to follow these stars because of their longer lifetimes.

\rev{Our model evolves the dust mass, but does not incorporate the grain size distribution or its evolution due to computational cost. While most semi-analytical models neglect this as well, detailed investigation of the grain size distribution has been done in a number of works \citep[e.g.,][]{Asano13b,Aoyama18,Hirashita19,Aoyama20}. These works find that the distribution is mostly unchanged within the first $\sim$ Gyr of age, which is older than the galaxies of our interest. The grain distribution can potentially affect the growth rate \revrev{and the opacity} to UV/optical radiation. For the former, this uncertainty is already absorbed in the parameter $t_{\rm acc,\odot}$ in equation (\ref{eq:t_grow}), which we find to affect the dust mass \revrev{by a factor $\lesssim 5$} (Figure \ref{fig:dust_z7}). For the latter, variations in dust distribution or composition can change the UV opacity by up to a factor $2$ from our assumption of $\kappa_{\rm d}=5\times 10^4\ {\rm cm^2\ g^{-1}}$ \citep{Hirashita19,Inoue20}. While the latter may influence the detectability of some high-redshift galaxies in the UV, it would not have a large impact on our conclusion on relating the dust ejection process and the dust mass, which is the main focus of this paper.}

As assumed for gas in \asloth, our model of dust assumes uniform density and spherical symmetry, and neglect the inhomogeneity of the dust distribution and/or the spatial displacement between stars and dust. An inhomogeneous distribution, such as a clumpy structure, may enable partial escape of photons and reduce the radiative acceleration of dust. Such possibilities were also raised in \cite{Ziparo23} to explain the blue colors of JWST galaxies, though the stringent constraints of the dust mass by ALMA \citep{Yoon22,Bakx23} would likely still require dust ejection to operate. 

We also have assumed perfect coupling between gas and dust, based on a heuristic timescale argument. However, recent studies point out that such dust-gas outflow is unstable to various classes of instabilities, which may decouple the dynamics of these two \citep{Squire18,Hopkins18,Hopkins22}. In this case radiation pressure may preferentially expel the dust, and hence the requirements for a recent extreme starburst for the JWST galaxies may be alleviated.

\rev{We further comment that the two assumptions in this work for the dust extent in the cold ISM are two extremes, in the sense that dust is exclusively located in either the compact stellar region where dust forms or the more diffuse ISM region. We expect that dust would coexist in both regions, and likely with different temperatures.} This is because acceleration becomes ineffective at \rev{the place} where the expanding dust becomes optically thin, which can happen within the extent of cold gas $R_{\rm s}$. Since we resolve the individual dust production by SNe, it may be possible to also evolve the radial distribution of dust by following its dynamical evolution under e.g. radiation pressure, gravity, and deceleration by sweeping ambient gas \citep[e.g.,][]{Murray05, Thompson15}, which may lead to a more realistic estimation of the attenuation and thermal emission of dust. We leave such attempts to a future study.

For models of efficient radiation feedback, one may worry that the ejection of gas can affect the parameter calibrations previously done in \asloth. The calibrations that are likely affected would be the star formation history and the reionization history. However, this feedback should affect only a small fraction of galaxies that are on the bright end in terms of the luminosity function, which are probably the case for JWST galaxies \citep[e.g.,][]{Mason23}. As star formation and reionization are expected to be dominated by the more numerous dimmer galaxies, we expect that the change to the global star formation at $z\lesssim 10$ used for calibration should be minimal. 

Nonetheless, depending on how rare these galaxies are, the existence of bursty star formation with possibly inefficient feedback may influence the detectability of galaxies at these redshifts \citep{Sun23,Shen23,Pallottini23} or even earlier ones \citep{riaz22}, and the explosions of stars that comprise them \citep[e.g.,][]{Tanaka13,Desouza14}. Future JWST observations in the years to come, along with simulations like done in our work, would be able to clarify the rich history of star-formation in the early universe.

\section*{Acknowledgements}
The authors thank Shafqat Riaz for providing us code for parameter scanning, and the anonymous referee for constructive comments that greatly improved the manuscript. We also thank Philip Hopkins, Seiji Fujimoto and Yoshinobu Fudamoto for commnets on dust physics. DT is supported by the Sherman Fairchild Postdoctoral Fellowship at the California Institute of Technology. YN is supported by JSPS KAKENHI grant No. 23KJ0728 and by the JSR Fellowship. TH acknowledges financial support from JSPS (KAKENHI Grant Numbers 19K23437 and 20K14464) and the German Environment Agency.

 \section*{Data Availability}
The modified version of the \asloth code used in this work is publicly available in the following github repository: \verb|https://gitlab.com/thartwig/asloth|. All data underlying this article are available on reasonable request to the corresponding author.



\bibliographystyle{mnras}
\bibliography{reference} 



\appendix

\section{Comparison with Dust Upper Limits from ALMA Galaxies}
\label{sec:alma_upper_limits}
\rev{The comparison of our model galaxies at redshift 7 with the observations by ALMA were restricted to sources that had detection of dust continuum in the rest-frame far infrared \citep[among other selection criteria;][]{Sommovigo22}, as they realize more robust measurements of the dust temperature. It is important to confirm that our model does not contradict with galaxies with non-detections of dust continuum, which is comparable in number to ones with detections \citep{Inami22}.}

\rev{Here we compare our models to the upper limits on the dust mass inferred for those galaxies lacking detection in the rest-frame far infrared. Inspecting the samples in \cite{Inami22}, 23 galaxies in the redshift range of $6.5<z<7.5$ have upper limits in the IR emission $L_{\rm IR}$ in the range $L_{\rm IR}\lesssim (3$--$6)\times 10^{11}L_\odot$. We use a formula $L_{\rm IR}=L_\odot(M_{\rm dust}/M_\odot)(T_{\rm dust}/8.5\ {\rm K})^{4+\beta}$ \citep{Sommovigo22} to relate $L_{\rm IR}$ and $M_{\rm dust}$. We adopt a dust emissivity index of $\beta=2.03$ following \cite{Sommovigo22}, and a dust temperature $T_{\rm dust}=40$ K that is in the lower end inferred from the samples in \cite{Sommovigo22}. We can then obtain a conservative upper limit on $M_{\rm dust}$ from the equation
\begin{eqnarray}
    M_{\rm dust} \lesssim  2.6\times 10^7M_\odot\left(\frac{L_{\rm IR}}{3\times 10^{11}L_\odot}\right) \left(\frac{T_{\rm dust}}{40\ {\rm K}}\right)^{-6.03}.
\end{eqnarray}
The estimated stellar masses for these galaxy samples are mostly in the range $4\times 10^{8}M_\odot<M_*<1\times 10^{10}M_\odot$ \citep{Bouwens22}. For our model galaxies in this stellar mass range, the dust masses are in the range $10^6M_\odot \lesssim M_{\rm dust}\lesssim 2\times 10^7M_\odot$ for $t_{\rm acc,\odot}=30$ Myr and $3\times 10^6M_\odot \lesssim M_{\rm dust}\lesssim 10^8M_\odot$ for $t_{\rm acc,\odot}=3$ Myr (Figure \ref{fig:dust_z7}). Thus the non-detections of dust for these galaxies are consistent with our models, and those with the largest stellar masses favor the model with $t_{\rm acc,\odot}=30$ Myr over $t_{\rm acc,\odot}=3$ Myr.}

\rev{We note that the upper limits sensitively depend on $T_{\rm dust}$, with the limit becoming a factor $\sim 3$ more stringent if we assume the average value $T_{\rm dust}=47$ K of the samples of \cite{Sommovigo22}. However it is possible that there is a selection bias on $T_{\rm dust}$ as well as $M_{\rm dust}$, with the IR-dim samples having systematically lower $T_{\rm dust}$ than the IR-detected samples.}


\bsp	
\label{lastpage}
\end{document}